\documentclass[sigconf]{acmart}
\renewcommand\footnotetextcopyrightpermission[1]{To appear in GECCO 2020} 
\usepackage{booktabs} 

\setcopyright{none}

\usepackage{enumitem} 
\setlist{leftmargin=6mm}
\usepackage{xspace} 
\usepackage{balance}
\usepackage{comment}

\usepackage{xcolor}
\usepackage{multirow}
\usepackage{graphicx}

\usepackage[algoruled, lined, linesnumbered]{algorithm2e}
\usepackage[textsize=tiny]{todonotes}
\usepackage{tikz}
\usepackage{pgfplots}
\usepackage{pgfplotstable}
\usetikzlibrary{arrows}
\usepackage[normalem]{ulem}
\usepackage{lipsum}
\usepackage{multicol}

\newcommand{\todoInline}[1]{\textcolor{red}{#1}\xspace}
\newcommand{\mahmoud}[1]{\textcolor{red}{\textbf{Mb:} #1}\xspace}
\newcommand{\Brad}[1]{\textcolor{red}{\textbf{Ba:} #1}\xspace}
\newcommand{\m}[1]{\textcolor{red}{\textbf{MW:} #1}\xspace}
\newcommand{\ignore}[1]{}

\newcommand{\approachOne}{\textsc{Approach~1}\xspace}
\newcommand{\approachTwo}{\textsc{Approach~2}\xspace}
\newcommand{\approachThree}{\textsc{Approach~3}\xspace}
\newcommand{\approachFour}{\textsc{Approach~4}\xspace}
\newcommand{\approachRRR}{\textsc{R3-validation}\xspace}

\usepackage{subcaption,caption,epstopdf,url,amsmath}

\usepackage[autolanguage]{numprint} 

\copyrightyear{2020} 
\acmYear{2020} 
\setcopyright{acmcopyright}\acmConference[GECCO '20]{Genetic and Evolutionary Computation Conference}{July 8--12, 2020}{Cancún, Mexico}
\acmBooktitle{Genetic and Evolutionary Computation Conference (GECCO '20), July 8--12, 2020, Cancún, Mexico}
\acmPrice{15.00}
\acmDOI{10.1145/3377930.3390245}
\acmISBN{978-1-4503-7128-5/20/07}

\begin{document}
\title{Towards Rigorous Validation of Energy Optimisation Experiments}

\author{Mahmoud A. Bokhari}
\affiliation{\institution{Optimisation and Logistics, School of Computer Science, University of Adelaide, Australia}
}
\affiliation{\institution{Computer Science Department, Taibah University, Kingdom of Saudi Arabia}}
\email{mahmoud.bokhari@adelaide.edu.au}

\author{Brad Alexander, Markus Wagner}
\affiliation{\institution{Optimisation and Logistics, School of Computer Science, University of Adelaide, Australia}}
\email{bradley.alexander@adelaide.edu.au} \email{markus.wagner@adelaide.edu.au}

\sloppy

\begin{abstract}
The optimisation of software energy consumption is of growing importance across all scales of modern computing, i.e., from embedded systems to data-centres. Practitioners in the field of Search-Based Software Engineering and Genetic Improvement of Software acknowledge that optimising software energy consumption is difficult due to noisy and expensive fitness evaluations. However, it is apparent from results to date that more progress needs to be made in rigorously validating optimisation results. This problem is pressing because modern computing platforms have highly complex and variable behaviour with respect to energy consumption. To compare solutions fairly we propose in this paper a new validation approach called \approachRRR which exercises software variants in a rotated-round-robin order. Using a case study, we present an in-depth analysis of the impacts of changing system states on software energy usage, and we show how \approachRRR mitigates these. We compare it with current validation approaches across multiple devices and operating systems, and we show that it aligns better with actual platform behaviour.

\end{abstract}


\maketitle


\section{Introduction}

In many applications of evolutionary search the fitness functions used during search are noisy and approximate.
A key way to address these limitations is to add a validation stage to the end of the search process to more thoroughly evaluate and rank the best individuals produced by the initial search process~\cite{gi:minisat}. One field in which validation is of increasing importance is the application of evolutionary search to improve the non-functional properties of programs. 
Examples of such properties include program execution time and energy use. 
Observed measurements of such properties arise from an interaction between the target program and the time-variant system state of its host platform. Unfortunately, as platforms become more complex, heterogeneous, and adaptive, it becomes much more difficult to model and compensate for changing system states. In turn, these complex interactions make it all too easy for validation runs of program variants to give misleading results. 
 
In this work we explore the problem of validating solutions from evolutionary processes aimed at optimising energy consumption on mobile platforms. This application domain is especially challenging for program optimisation, due to factors including: diversity of platforms, system configurations, and highly complex and adaptive system states. We show how these system features can combine to thwart current published approaches to validation runs of programs evolved for improved energy and execution speed. 
We do this by first characterising the time-variant effect of system state on a program run on multiple platforms. We show how, in spite of reasonable efforts to achieve uniformity of system state, successive experiments exhibit large and unpredictable variations in key system features. Finally, we propose a simple approach to the scheduling of the running of program variants, called rotated-round-robin (\approachRRR) that increases the chance of variants being run in the same {\em{set}} of system states. We measure how effective it is in allowing variants to sample similar system states and, thus, enabling more specificity and sensitivity in comparisons of software variants. 
 
The remainder of this article, we first review literature related to system states and to validation approaches in Section~\ref{sec:literature}. 
Then, we provide a horrifying motivating example in Section~\ref{sec:horror}. 
To deal with these, we define a number of validation approaches in Section~\ref{sec:approaches}, including our \approachRRR.
Using a case study in Section~\ref{sec:casestudy}, we observe and characterise system states, and we demonstrate how \approachRRR mitigates the variations. 
In Sections~\ref{sec:specificity} and~\ref{sec:comparison}, we compare the validation approaches in terms of sensitivity and specificity across a number of different devices and operating system versions. 
Finally, we wrap up in Section~\ref{sec:conclusion}.

\section{Literature Review}
\label{sec:literature} 

\paragraph{Impact of System State on Non-Functional Properties}
The impact of highly-variable system state on measured program performance has been described in many works in the literature. Program execution time can be greatly affected by the state of cache memory~\cite{variabilityCPU}; cache miss-rates; context switches~\cite{threadContextSwitch}; JIT compilation settings~\cite{JvmNonDeterminism1, JvmNonDeterminism2, JvmNonDeterminism3}; memory layout~\cite{stabilizer}; garbage collection policies~\cite{JvmNonDeterminism3, JvmNonDeterminism4}; and OS environment variables~\cite{producingWrongData}.
Even when a platform is rebooted between runs there can be significant variation in benchmark runtimes~\cite{InitialSystemState}.
Furthermore, the works in~\cite{benchmarkPrecision, rigorousSystemStates} show that measurements of the benchmark operation are subject to inherent and external random changes to the benchmark operation, as well as the system state prior to execution. In this work we show that changes in system states between and during validation runs on mobile platforms can also impact energy measurements.
This variation in system state can affect the sensitivity and specificity of conclusions drawn from validation runs.
\paragraph{Current Validation Approaches}
Much existing work on using search to optimise non-functional properties of programs is described in~\cite{petke:survey}. We briefly summarise the approaches to validation described in this work and in more recent related publications. It should be noted that not all works included a validation stage to confirm the performance of solutions. Where validation was done it involved re-sampling individual program variants. 

In work optimising energy use on mobile platforms, Schulte et al.~\cite{Schulte:postCompilerOptimisation} and Li et al.~\cite{oledWebApps} used external meters with constant voltage. While external meters are precise they fail to capture different voltage states that applications encounter (and, potentially, trigger) when running on battery. This is significant because changes in voltage have an impact on energy use~\cite{energyConsumptionBook, Vedge}. Bokhari et al.~\cite{Bokhari2018mobiquitous} avoided this issue by using devices' internal meters, at the cost of noisier energy measurements~\cite{Bokhari2017arxivValidation}.

An alternative approach to meters is to use energy models based on hardware (HW) counters~\cite{nathan:guavaOptimisation,gi:minisat}. Although HW models produce cleaner signals, they are affected by the current system state as well as non-determinism in the virtual machine behaviour caused by factors such as garbage collection, thread scheduling and Just-In-Time optimisation~\cite{blackburn2016truth}. In addition, 
such models abstract away at least some of the relevant environmental conditions and only capture the assumptions on which they were built. 

Other work that relied on energy models has used simulation~\cite{davidThesis} or hybrid linear models based on small-scale energy sampling~\cite{Linares-Vasquez:gemma}. As with HW counter based models these models abstract away from at least some important details of system state.

There are also works that
use the unmodified target platform for validation
tests~\cite{nathan:guavaOptimisation,gi:minisat,Schulte:postCompilerOptimisation,oledWebApps,Bokhari2018mobiquitous}. 
These works conducted multiple performance measurements and carried out non-parametric statistical tests and effect size measurements to achieve some confidence in the observations. 

One of these works, \cite{gi:minisat}, revealed individuals that performed well only during optimisation -- thus showing the potential value of validation. Other publications had less complete data with \cite{oledWebApps} reporting only the average of 10 samples. Likewise, the work in \cite{Schulte:postCompilerOptimisation} reported the \textit{p-values} of comparisons without naming the statistical test used and did not report the effect sizes from this test. Like most prior work we equip our \approachRRR approach with non-parametric statistical analysis and effect size measure to compare between obtained solutions' samples.

Although most prior work applied rigorous statistical analyses, most work has not explicitly 
acknowledged or mitigated changing system states (see, e.g., \cite{gi:minisat,Schulte:postCompilerOptimisation,oledWebApps,Bokhari2018mobiquitous,davidThesis,Linares-Vasquez:gemma}). 
In an attempt to consider these, Nathan et al.~\cite{nathan:guavaOptimisation} avoided some aspects of system noise by turning off the JIT and garbage collection Java systems whilst optimising for execution speed and reactivated these during validation runs. However, in both cases the execution times were modelled by virtual machine instruction counts which are somewhat decoupled from wall-clock time measurements.

A different approach to controlling the system was taken by Bruce et al.~\cite{bruce:energy}, who suggested running all validation
experiments immediately after a system reboot to avoid the random initial state of the system. Unfortunately, on mobile platforms this protocol doesn't guarantee similar states, partly because, not all system services start immediately after reboot but also because the battery state of the system can trigger changes in system behaviour such as suspending background processes.

In summary, we propose that more progress needs to be made in rigorously validating optimisation results that come from running software on modern computing platforms. 
Our proposed \approachRRR approach mitigates these issues by scheduling solution executions in similar system conditions, that allowing a fair comparison to be conducted. 


\section{A horrifying motivating example}
\label{sec:horror}

We now illustrate how temporal changes in system state on Android platforms are reflected in measurements of energy consumption of even only a single program variant. 
We show that this changing system state can make a variant appear significantly different from itself even if the measurements take place between system reboots. 

But first, we introduce the software and hardware used throughout this article.

\subsection{Target software and hardware} 

In their optimisation research Bokhari et al.~\cite{Bokhari:geccoGi2017,Bokhari2018mobiquitous} targeted Facebook's \emph{Rebound}, which is an open-source  physics library modelling spring dynamics. Rebound is used by popular Android applications such as Evernote, Slingshot, LinkedIn and Facebook. The authors used multi-objective optimisation to explore trade-offs between energy consumption and physical accuracy. In particular, \cite{Bokhari2018mobiquitous} explored the use of the direct energy measurements (named "raw") and also the use of models based on jiffies (a time unit on Android phones) and executed lines-of-code (LOC). 
The experiment that used the raw energy measurements resulted in a Pareto front containing 10 solution variants named {\em{raw1}} to {\em{raw10}}. In the following, we show how system state affects the energy consumption of these variants and how this can impact on validation.

The target devices for our experiments are the HTC Nexus 9 and the Motorola Nexus 6. Both are equipped with the Maxim MAX17050 fuel gauge chip that compensates measurements for temperature, battery age and load~\cite{maxim}. We use this internal chip for all validation experiments, as other work has shown a good correlation of the measurements with those performed with an external meter~\cite{Bokhari2017arxivValidation}. 

We also use the Motorola G4 Play with Android 6 and 7, and the Sony XZ with Android 8.0 and 8.1. As these are not equipped with comparable battery fuel gauge chips, the OS provides estimates based on voltage readings and course-grained internal power profiles.

\subsection{Observations from a simple experiment} \label{singleExpOBD}

\begin{figure}[t]
\centering

\includegraphics[width=\linewidth]{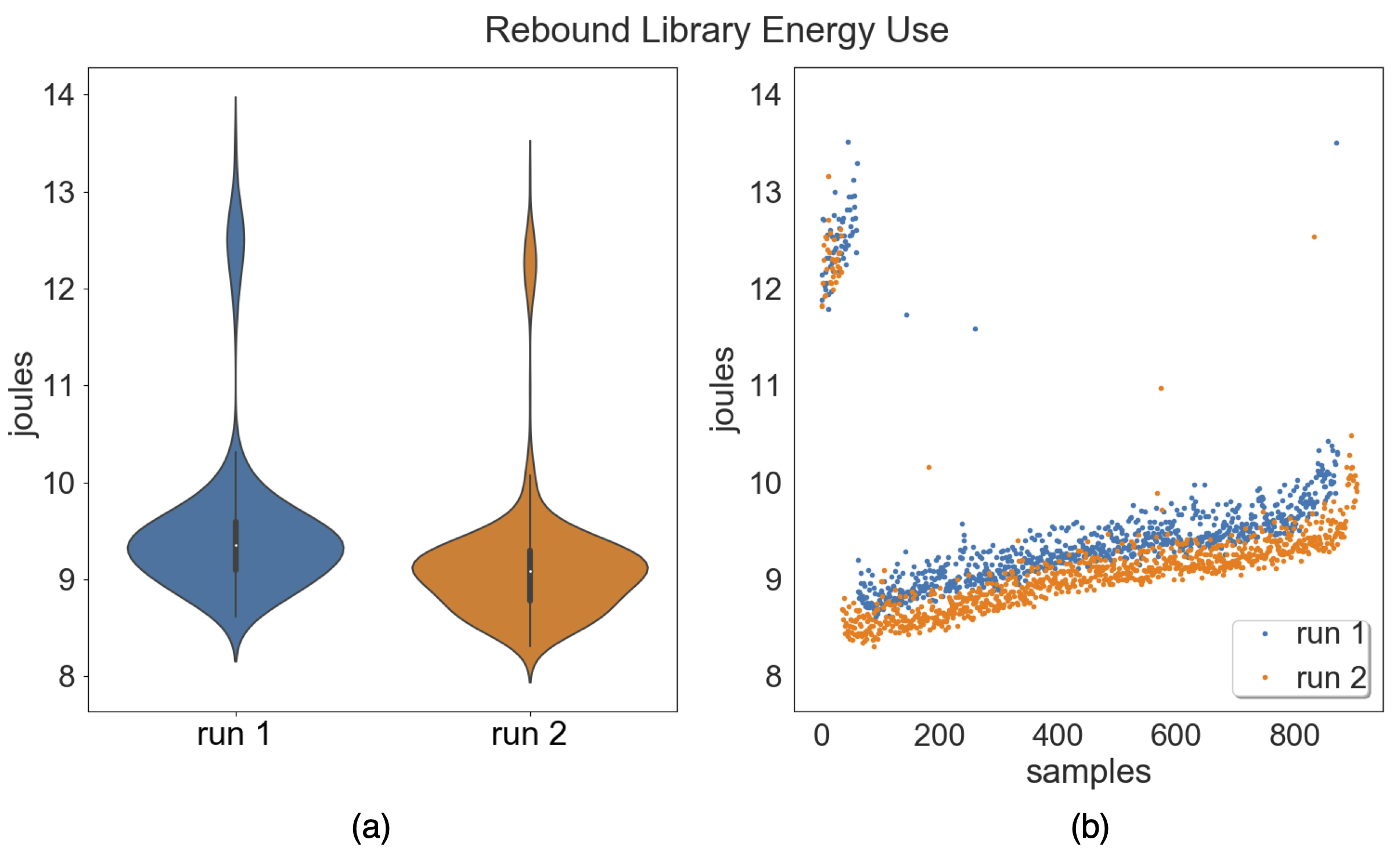}
\caption{Energy use of the Rebound library in two experiments. Violin plots (a) and sample distributions (b). The device was rebooted and recharged between the two experiments.}
\label{fig:violinPlot2originalConfigRunsCEC-Nexus6}
\end{figure}

To motivate the need for rigorous approaches to validation, we start by analysing the results of a simple experiment: we repeatedly run the original Rebound library, then reboot and recharge the device, and then repeatedly run the original Rebound library again. We do this on a Nexus 6 using Android 7.

Figure~\ref{fig:violinPlot2originalConfigRunsCEC-Nexus6}(a) shows the distribution of the energy consumption, which is clearly multimodal. Moreover, the energy consumption from the second run is skewed significantly lower -- this means that in this experimental setting the variant is demonstrated to be significantly different from itself! This indicates that this -- admittedly very simple -- validation process lacks statistical specificity. 
Figure~\ref{fig:violinPlot2originalConfigRunsCEC-Nexus6}(b) provides a deeper look into the individual runs of the two experiments. In general, the measurements exhibit significant random and systematic noise. At the beginning of the experiments, the consumption is considerably higher than during the rest of the experiments. Moreover, the first experiment is clearly consuming more energy than the second experiment in most of the runs. 

Although the approach of rebooting and recharging seems to be prudent at first glance, rebooting a device can impact the energy use of the subsequent runs. This introduces clear challenges for researchers attempting to validate the outcomes of energy optimisation results, as software variants need to run under the same conditions to be compared fairly.

\begin{figure}[t]
\centering
\includegraphics[width=5cm]{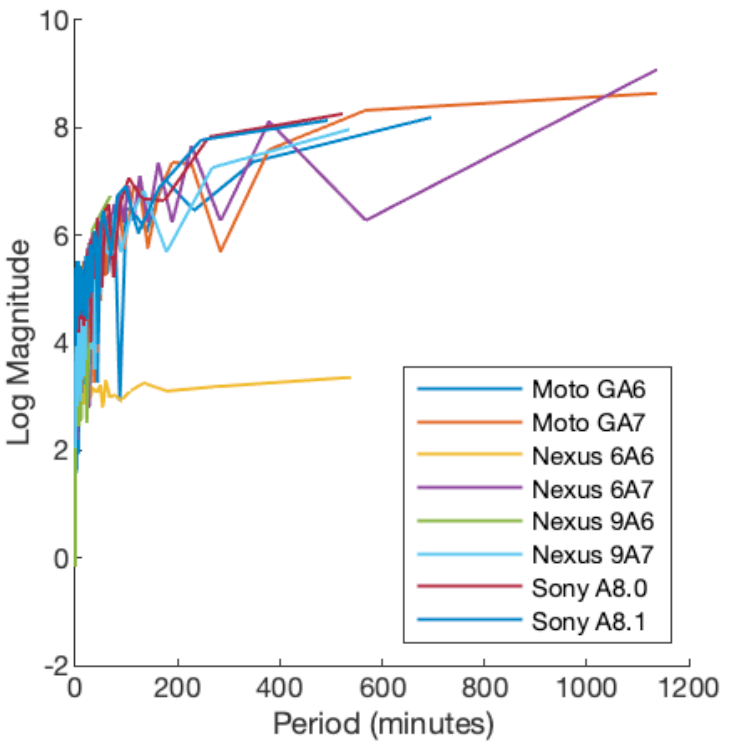}
\caption{Log-scale spectral analysis of energy
measurements of the original Rebound
library run repeatedly on a full charge on eight different platform variants. On all but one platform long period changes in measurement environment dominate. 
}
\label{fig:periodicity}
\end{figure}

Long-frequency changes in system state (evident in Figure~\ref{fig:violinPlot2originalConfigRunsCEC-Nexus6}(b)) are large enough to impact specificity even when validation runs are made within a single charge of a device. Such long-wave changes can be observed on multiple platforms. 
Figure~\ref{fig:periodicity} shows a spectral analysis of the measured energy consumption of the original rebound library on eight different hardware/OS combinations. These curves show that measurement variability induced by long-period changes in system state dominate on all platforms except Nexus 6 - Android 6 (which exhibits globally low system-induced variation). These large signals at low-frequency show that is problematic to make direct comparisons of software variants run a long time apart. Conversely, the magnitude of variation is moderate for variants run within 50 minutes of each other. That is, we are more likely to capture similar system states if we run variants close together in time. This observation can help inform the design of our validation approach.


\section{Validation Approaches}
\label{sec:approaches}

In this section we describe a number of  
alternative approaches to scheduling
runs of program variants for validation. 
In all approaches we assume that there are $n$ required samples of each program variant, where we label variants $A,B,C,D,\ldots$. We assume that we are re-sampling runs for long enough so that the non-functional property is detectable above the noise on the internal meter of the device. In all of our experiments involving the aforementioned Rebound library and its configurations, 
measuring the total energy consumption of a configuration once 
takes between 20 seconds and a minute depending on the platform.  

The first approach is called \approachOne and is the modal approach used for validating the outcomes of the optimisation experiments in the literature \cite{Schulte:postCompilerOptimisation,oledWebApps,nathan:guavaOptimisation,gi:minisat}. This approach starts with a {\em{setup}} phase where the device is rebooted and initialised by disabling unused hardware components after a recharge. 

{\em{Setup}} is followed by the experimental phase where each variant is run $n$ times in turn. So for example, if we had variants $A,B,C,D$ and $n=4$ then \approachOne would produce the schedule: $\mbox{\em{setup}},AAAA,BBBB,CCCC,DDDD$.

The second approach, \approachTwo, re-samples one solution the required number of times starting from a full battery. The device is then {\em{setup}} again and then the second solution is run and so on. Under this approach with $n=4$ the variants above would run with the order: $\mbox{\em{setup}},AAAA,\mbox{\em{setup}},BBBB,\mbox{\em{setup}},CCCC,\mbox{\em{setup}},DDDD$. 

A third approach, \approachThree, 
interleaves the samples. So, for $n=4$ and the variants above we would have the schedule: $\mbox{\em{setup}},ABCD,\mbox{\em{setup}},ABCD,\mbox{\em{setup}},ABCD,\mbox{\em{setup}},ABCD$. 
A-priori, as long as each round $ABCD$ happens sufficiently quickly, then the variants in each round have a better chance of running in a similar system state than the previous approaches\footnote{The spectral analysis in Figure~\ref{fig:periodicity} indicates similarity could be achieved with rounds taking less than 20 minutes.}.  
A fourth approach, \approachFour is similar to \approachThree but runs all trials on a single charge. So for, $n=4$ \approachFour would produce the schedule: $\mbox{\em{setup}},ABCD,ABCD,ABCD,ABCD$. This approach was used in \cite{bruce:energy}.

Our final approach \approachRRR
is similar to \approachThree above except that the variants are left-shifted between setup phases and up to $\pi$ sets of trials are run on a single discharge. 

Thus for $n=8$ and $\pi=2$ we would have:

\noindent\hspace{6mm}\mbox{\em{setup}},ABCD,ABCD,\mbox{\em{setup}},BCDA,BCDA,

\noindent\hspace{6mm}\mbox{\em{setup}},CDAB,CDAB,\mbox{\em{setup}},DABC,DABC

\vspace{2mm}
The value of $\pi$ is chosen so as to not
deplete the battery below 20\% in a discharge cycle.\footnote{In our experiments system noise increases greatly if experiments proceed below 20\% charge.} This approach is described in Algorithm~\ref{alg:r3}.

\begin{algorithm}
\SetKwFor{RepTimes}{repeat}{times}{end}
\DontPrintSemicolon
\SetAlgoLined

\SetKwInOut{Input}{input}

\Input{Given a set $S$ of $N$ configurations, and 
$\pi$ the number of times a permutation is to be repeated within a discharge cycle.}

Create a permutation $\Pi$ based on $S$.\;

\RepTimes{$N$} {

  Reboot and recharge the device.\;
  
  \RepTimes{$\pi$} {
    Execute permutation $\Pi$ on the device.
  }
  Rotate $\Pi$ left by one position.\;
 }
\caption{Round-Robin \& Rotate (\approachRRR) validation protocol}
\label{alg:r3}
\end{algorithm}

\paragraph{Claims} 
We claim that in the systems and software variants we review, \approachRRR is able to sample similar sets of system states across program variants. 
We also claim that, in our experiments, \approachRRR is better for validation than the four alternative approaches listed above in terms of both test specificity, test sensitivity. 
We address these claims in turn, starting with system states in the next section and addressing specificity in Section~\ref{sec:specificity}, and sensitivity in Section~\ref{sec:comparison}.

\section{Case Study - Characterising system states and mitigating them}
\label{sec:casestudy}

In this case study, we analyse data captured from a validation run using \approachRRR.
We then look back through the encountered states to show how \approachRRR these have been 
consistently sampled for each variant. 

As mentioned before, we use the variants raw1 to raw10 from~\cite{Bokhari2018mobiquitous} and the same platform configuration (Nexus 9 Android 6). Combining \approachRRR 
with a targeted number of samples of at least 30 means that we need to restart the device 11 times and between each restart we run each configuration (in a rotating fashion) 3 times; this results in a total of 33 samples for each configuration.

\begin{figure}[h]
\includegraphics[width=\linewidth]{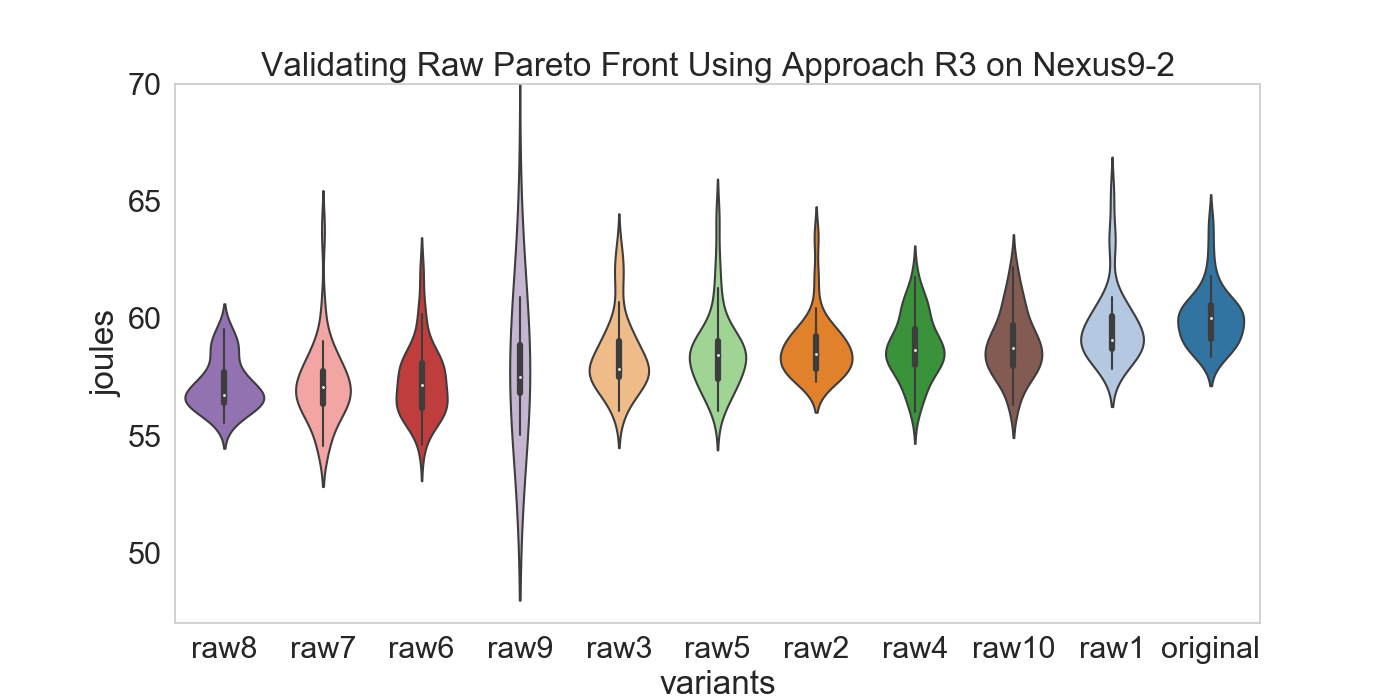}
\caption{Results of validating 11 configurations using \approachRRR. The configurations are ordered based on their median energy consumption.}
\label{fig:violinRawRoundRobinN9-2-M}
\end{figure}

Figure~\ref{fig:violinRawRoundRobinN9-2-M}
shows the distributions of energy measured by sampling using the \approachRRR sampling process. It can be seen that most distributions are compact and have a similar shape (slightly skewed to the right). The exception is the variant raw9 which has some outliers. The raw3 variant also has an extreme outlier (97 joules) which is not included in this plot. Note that the original ordering of the variants during optimisation was raw1, raw2, ... raw10, original so the validation approach has reordered some of the evolved variants.

\subsection{System State Analysis}
\approachRRR is designed to help mitigate the effect of changing system states when comparing program variants. In designing these protocols it is interesting to observe how system states change. Some  first insights were provided in Figures~\ref{fig:violinPlot2originalConfigRunsCEC-Nexus6} and~\ref{fig:periodicity}. However to see how deep and systematic changes in system state are, it is worth investigating these changing states in more detail. 

\begin{figure}
\includegraphics[width=\linewidth]{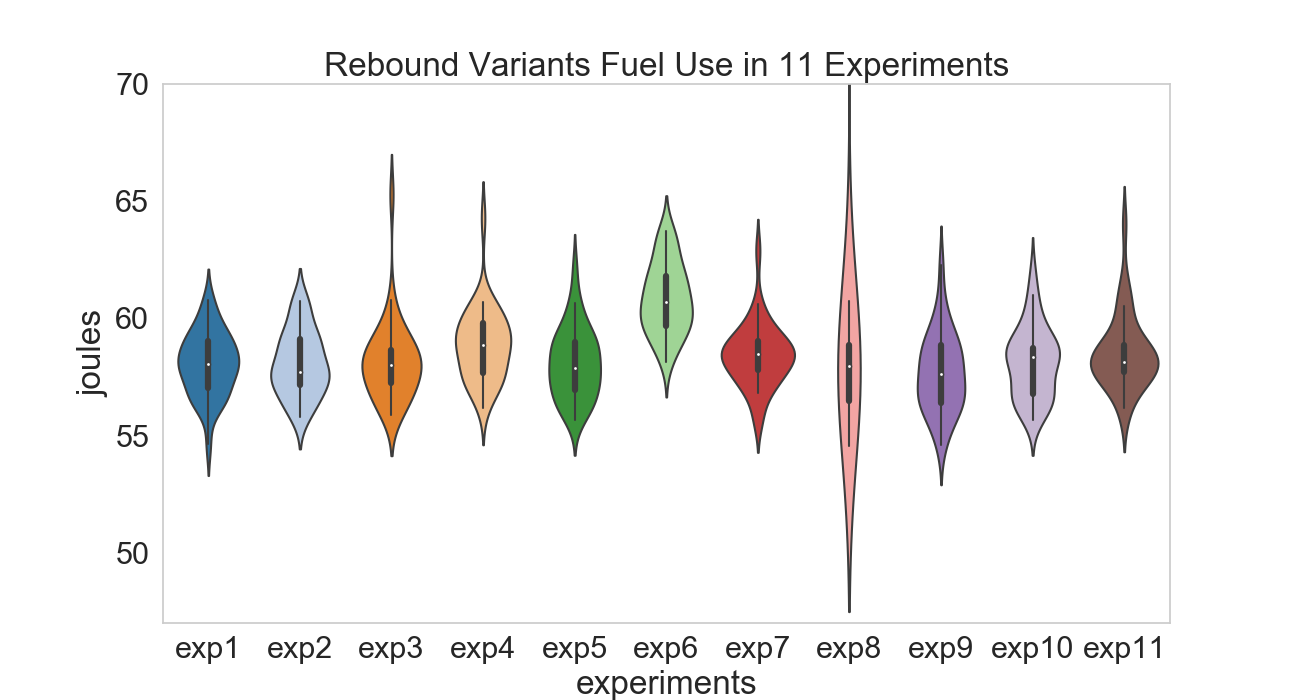}
\caption{Each distribution here is run of the variants original, raw1 ... raw10 executed three times. That is, the same workload is run for each of exp1 through to exp11. As can be seen there is considerable variation in distributions, which is attributable to changing system state.}

\label{violinfueluseReboundRawExperimentSamples-nexus9-2-M}
\end{figure}

In order to see the impact of changing system state in detail we ran an identical workload of the program variants original, raw1 ... raw10 executed, in round-robin fashion, three times. We 
then restarted the device and ran the experiment again and so on, 11 consecutive times on the Nexus 9 Android 6 platform. We name these 11 experiments exp1,...,exp11. For each experiment we sampled, energy and other information relating to system state. 

Although Rebound is a CPU-bound library that does not utilise other hardware components, system states impacted its energy consumption during the eleven round-robin experiments. The  plots in Figure~\ref{violinfueluseReboundRawExperimentSamples-nexus9-2-M} show the distribution of Rebound variants' energy use.

In general, each experiment, even though it was running an identical workload on a freshly booted machine produced a different distribution. The sixth experiment in particular used significantly more energy than the others and there seems to be a gentle upward trend in energy consumption overall.

\begin{figure*}
\centering
\begin{subfigure}{.33\textwidth}
  \centering
  \includegraphics[width=.99\linewidth]{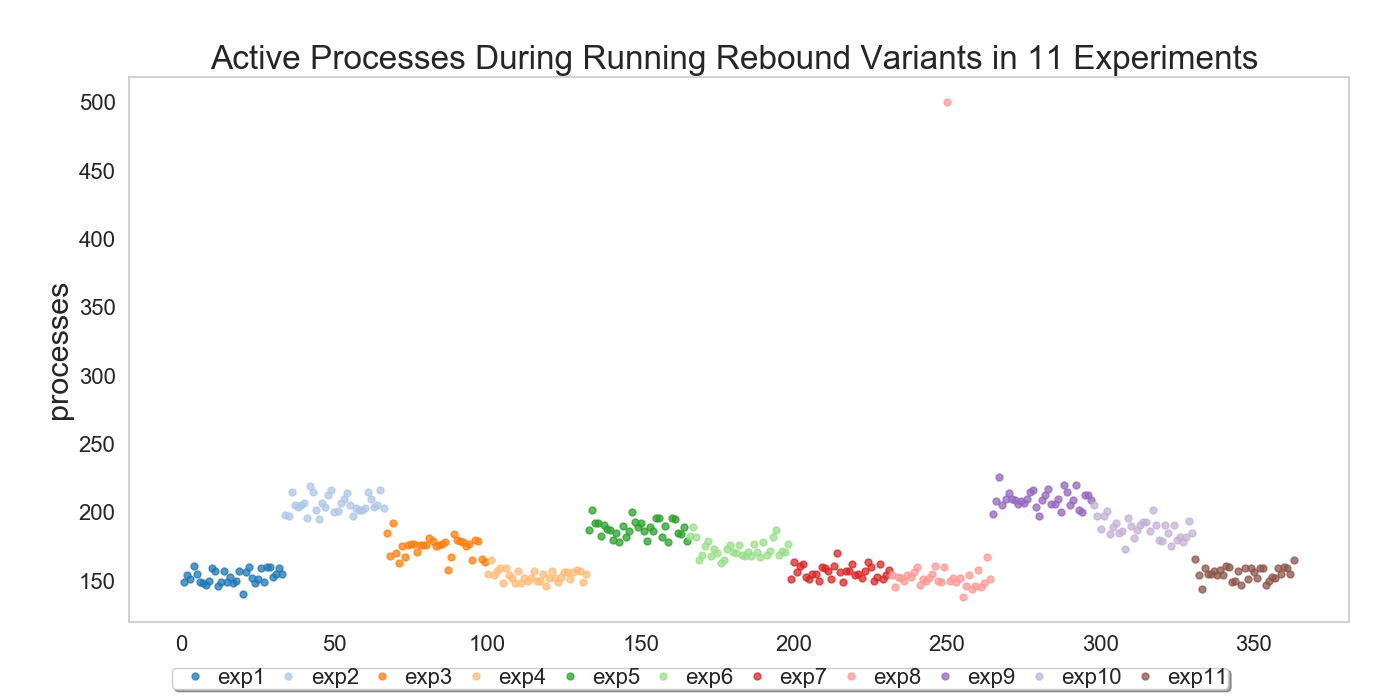}
  \caption{Active processes}
  \label{ActiveProcessesReboundRawExperimentSamples-nexus9-2-M}
\end{subfigure}%
\begin{subfigure}{.33\textwidth}
  \centering
  \includegraphics[width=.99\linewidth]{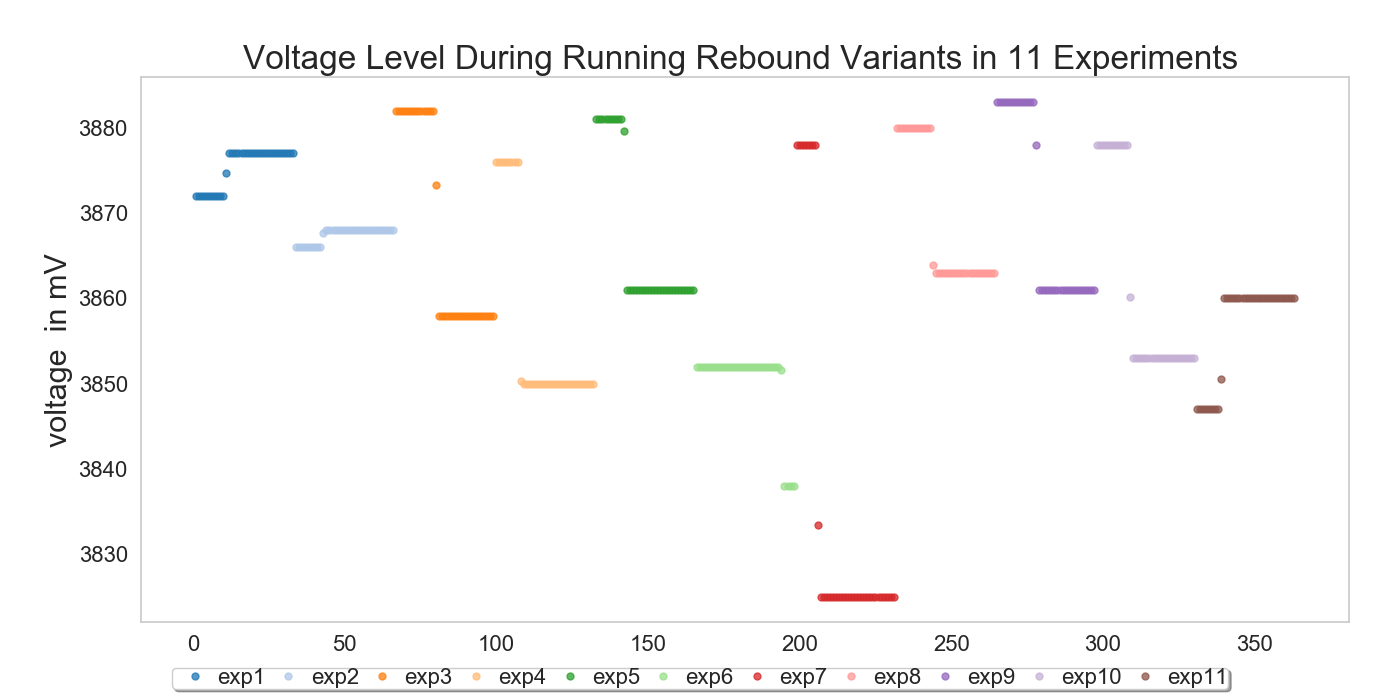}
  \caption{voltage}
  \label{VoltageLevelReboundRawExperimentSamples-nexus9-2-M}
\end{subfigure}
\begin{subfigure}{.33\textwidth}
  \centering
  \includegraphics[width=.99\linewidth]{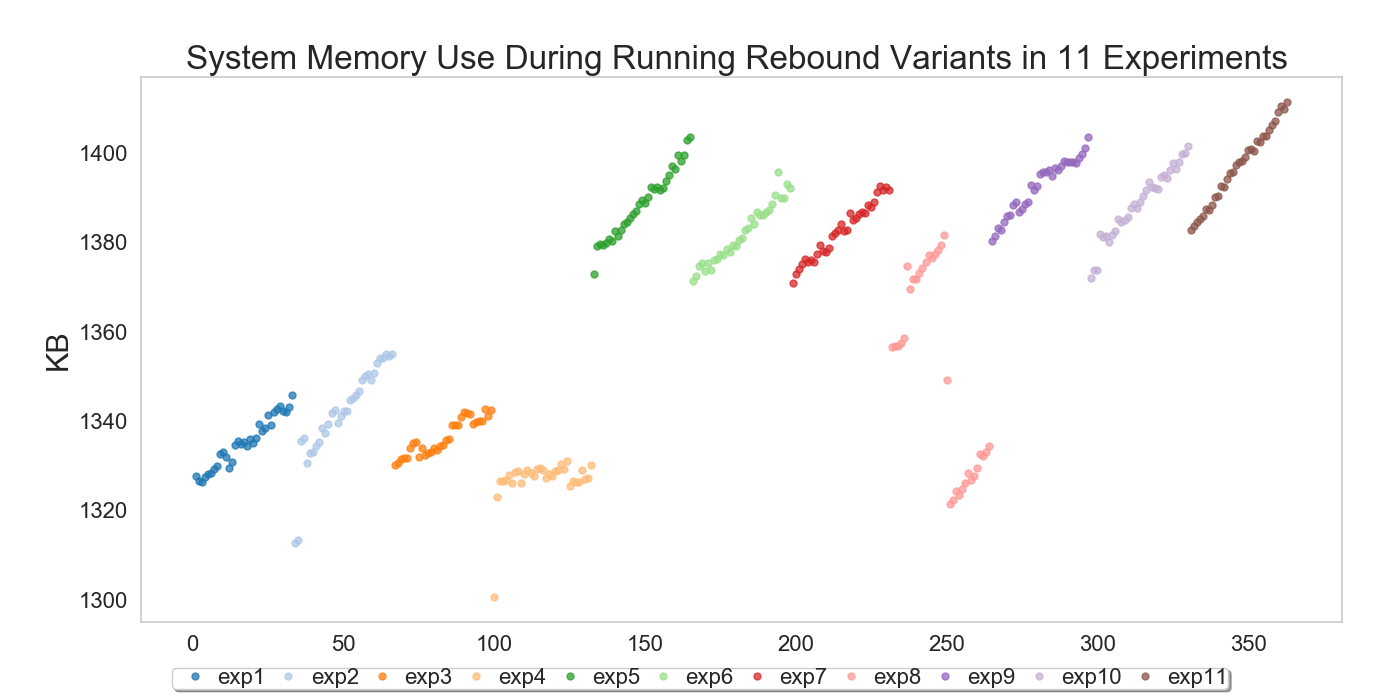}
  \caption{System memory use}
  \label{SystemMemoryUseReboundRawExperimentSamples-nexus9-2-M}
\end{subfigure}
\begin{subfigure}{.33\textwidth}
  \centering
  \includegraphics[width=.99\linewidth]{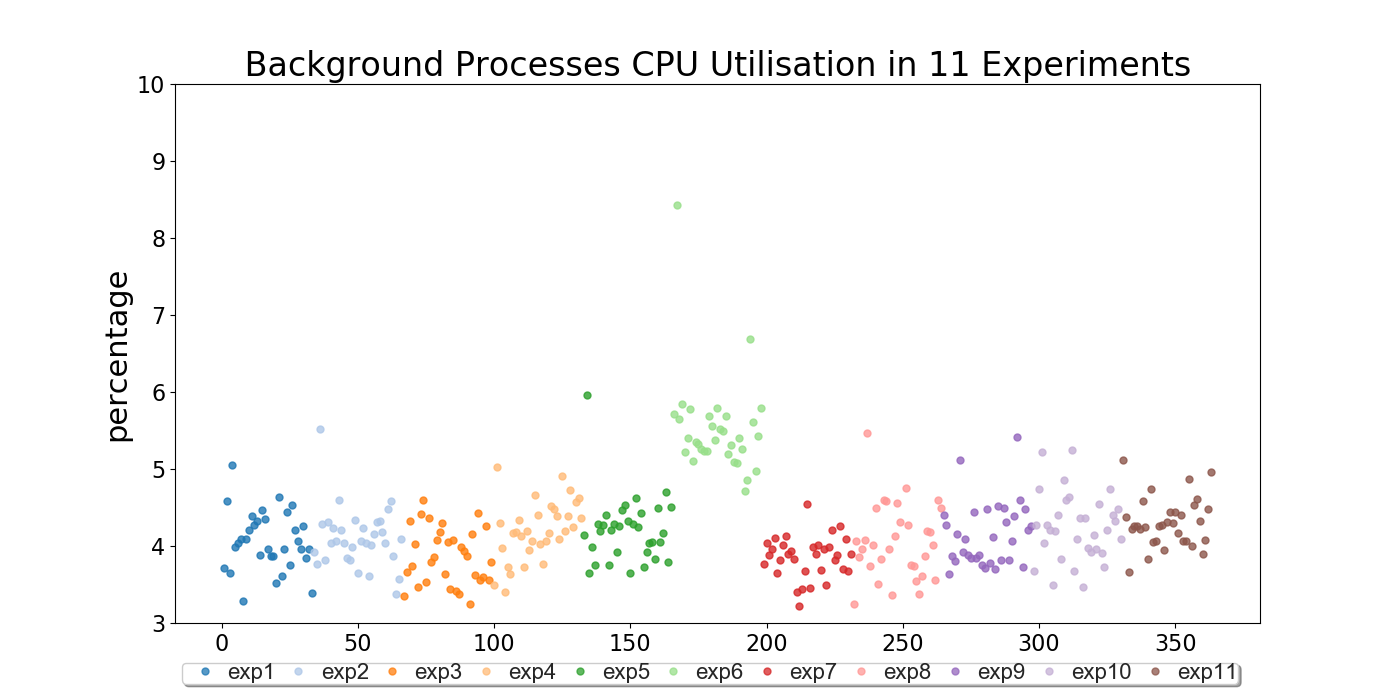}
  \caption{Background processes CPU utilisation}
  \label{bgCpuUtilisation-nexus9-2-M}
\end{subfigure}
\begin{subfigure}{.33\textwidth}
  \centering
  \includegraphics[width=.99\linewidth]{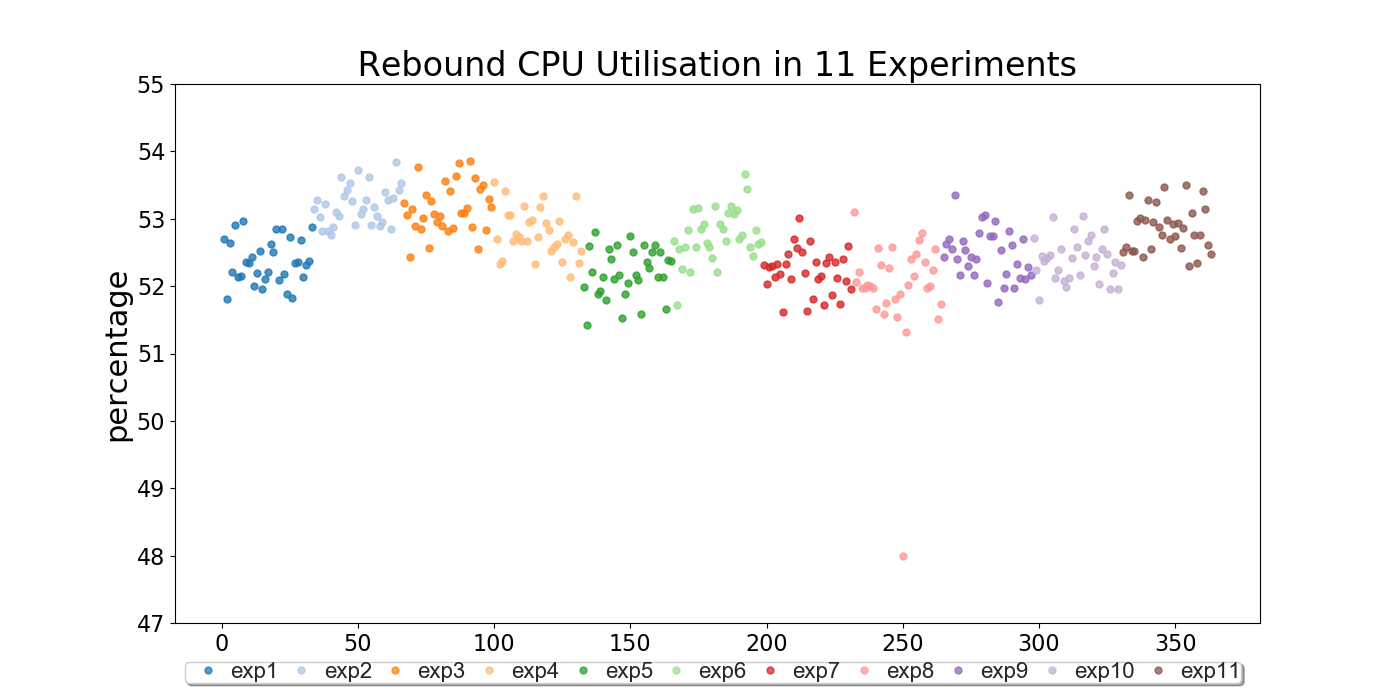}
  \caption{Rebound CPU utilisation}
  \label{reboundCpuUtilisation-nexus9-2-M}
\end{subfigure}
\begin{subfigure}{.33\textwidth}
  \centering
  \includegraphics[width=.99\linewidth]{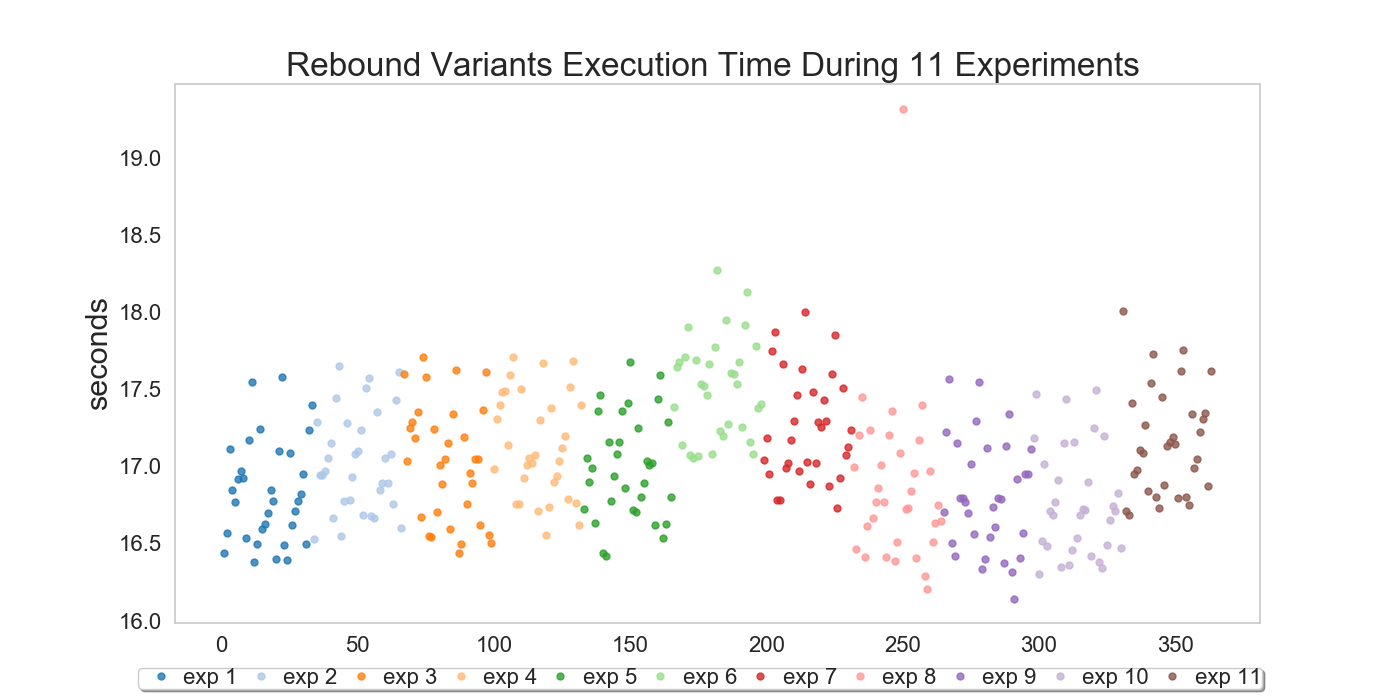}
  \caption{Rebound Execution Time}
  \label{reboundExecutionTime}
\end{subfigure}
\caption{Various measurements during the execution of Rebound variants in the 11 experiments. In each experiment 3 sequences of the solutions were exercised before recharging and rebooting the device. The difference between these experiments are the sequences of solutions (i.e., the permutations): the first experiment had original, raw1 ... raw10 executed 3 times}
\label{fig:raw11exps}
\end{figure*}

In the following, we shall dive a little deeper. 
Figure~\ref{ActiveProcessesReboundRawExperimentSamples-nexus9-2-M} shows individual samples of active background processes collected during the eleven experiments -- each experiment's dataset is shown in a different colour. In general, the overall trend fluctuates from experiment to experiment. The number of active processes ranges approximately from 140 to 250 processes during variants execution with one extreme outlier of 500 processes.
This outlier occurred during the execution of one variant of raw3 (which happened to register extreme energy use). Further investigation  of logs revealed that at that time Google Services' processes such as Google Play, Google Mobile Services and Google Cloud Messaging were, unsuccessfully,  trying to perform sync operations and pull updates\footnote{Note that all communication interfaces were turned off.}.

Another observation from Figure~\ref{ActiveProcessesReboundRawExperimentSamples-nexus9-2-M} is that samples collected in the same experiment are closer to each other in terms of active processes than other experiments. This indicates that, in terms of active processes, system state varies between system setups (reboot/recharging).

Voltage level can affect the amount of energy used by a program run and, ideally, it is good to run variants in a similar voltage state. 

Figure~\ref{VoltageLevelReboundRawExperimentSamples-nexus9-2-M} shows the voltage supply levels used during the round-robin experiments. Note that the initial voltage supply  differs slightly in each experiment even though we attempted to start each at the same battery level by recharging  between each experiment.
Furthermore, the trace of voltage levels varies drastically between experiments. For example, the seventh experiment (red dots) has a steeper drop in voltage levels compared to the rest. The second, third, ninth and tenth experiment had roughly the same drop in voltage (by about 24 mV).
Counter-intuitively three experiments (1,2 and 11) showed an increase in voltage. 
 This contrasts starkly with models in the  in the literature that assume a linear and an inverse correlation between voltage and state of discharge~\cite{Zhang:powerBooster,powerTutor}. 
It is worth mentioning that batteries are pre-configured and therefore experimenters can not change or control the voltage supply. Such extreme conditions show that no reasonable effort could guarantee the battery state to be the same for each experiment. It should be noted, however, that the battery state can stay steady for periods of time which means that two variants run close together in time have a chance of capturing a similar battery state.

Another aspect that cannot be fully controlled is the overall system memory consumption. This is due to the running background processes and the complex Android memory management system~\cite{highPerformanceAndroidApps}. Figure~\ref{SystemMemoryUseReboundRawExperimentSamples-nexus9-2-M} depicts interesting system memory usage behaviours. The memory consumption during the the eleven experiments are spread over different distributions. In addition, after the first four experiments, the usage suddenly increased by 5\% for the rest of the runs. As one might expect, the system memory fills up steadily as each experiment is executed -- except in the fourth experiment where it was stable and (again) in the eighth experiment that includes a large drop in memory consumption. No explanation was found for the stable memory consumption in exp4 but the drop in exp8 is due to a system update putting a demand on memory resources and thus invoking a garbage collection\footnote{To maximise responsiveness, Android reclaims memory from live and even dead processes only when it has to.}. This event also coincided with a peak in active processes.

As a consequence of the drastic increase in number of active processes demonstrated in Figure~\ref{ActiveProcessesReboundRawExperimentSamples-nexus9-2-M}, the CPU utilisation increased considerably to accommodate the extra workload. This is clearly depicted in Figure~\ref{bgCpuUtilisation-nexus9-2-M} which shows the CPU utilisation for the background processes. As can be seen, overall, the background processes occupied roughly 3\% to 8\% of the CPU in most of the time. However, in the eighth experiment where the device state changed to pulling updates and syncing states, their utilisation increased to about 40.5\%; the outliers were dropped from the plot to focus on the the other runs. 

Focusing just on Rebound's CPU use there was still variation between experiments.
Figures~\ref{reboundCpuUtilisation-nexus9-2-M} and~\ref{reboundExecutionTime} show the amount of CPU utilised and runtime needed by Rebound variants in the eleven experiments. Overall, the utilisation fluctuates between 48\% to 54\%, and the runtime varies between 16 seconds and 18.5 seconds. Interestingly, these two appear to be positively correlated, not negatively as one might expect for fixed loads. 

In summary, we have observed that the system states affected the energy consumption. In addition, our fine-grained investigations of the experiments show that no reasonable effort to control conditions could guarantee that each solution runs on the same system state. Thus, we need to cope with the  reality that the system state is will vary for each experiment run. Using our proposed method of validating solutions in a rotated-round-robin style, we attempt to ensure that such changes in state affect all solutions fairly. 

\subsection{Mitigating Changes in System State}
\begin{figure*}
\centering
\begin{subfigure}{.33\textwidth}
  \centering
  \includegraphics[width=.99\linewidth]{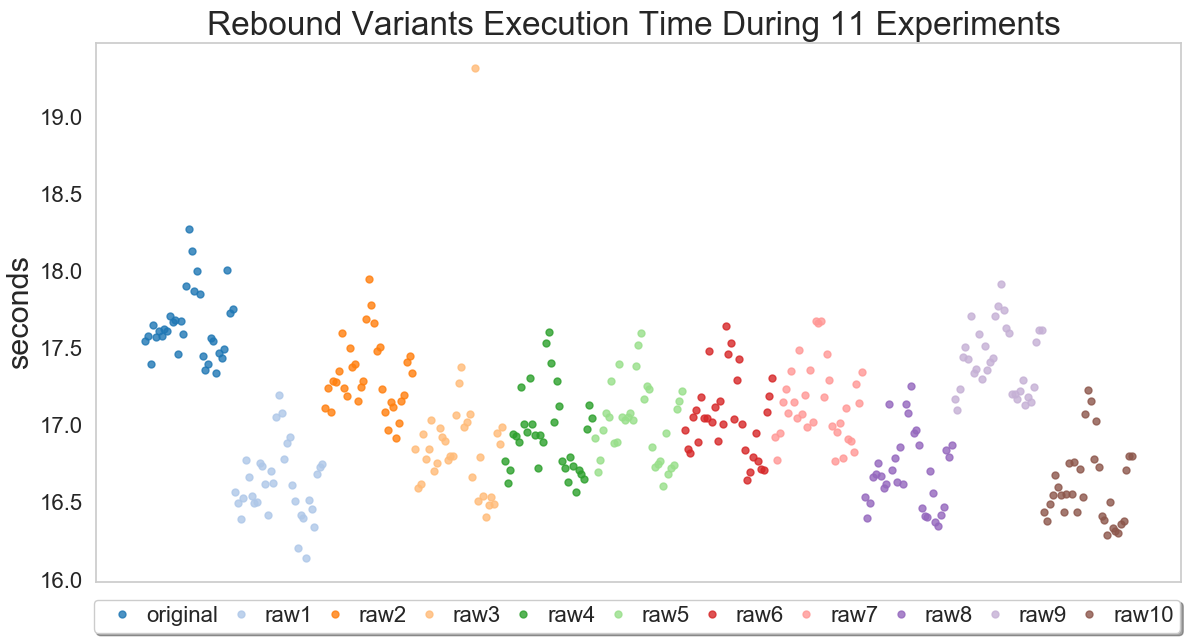}
  \caption{run-time}
  \label{scatterExecutionTimeReboundRawFronts-nexus9-2-M}
\end{subfigure}
\begin{subfigure}{.33\textwidth}
  \centering
  \includegraphics[width=.99\linewidth]{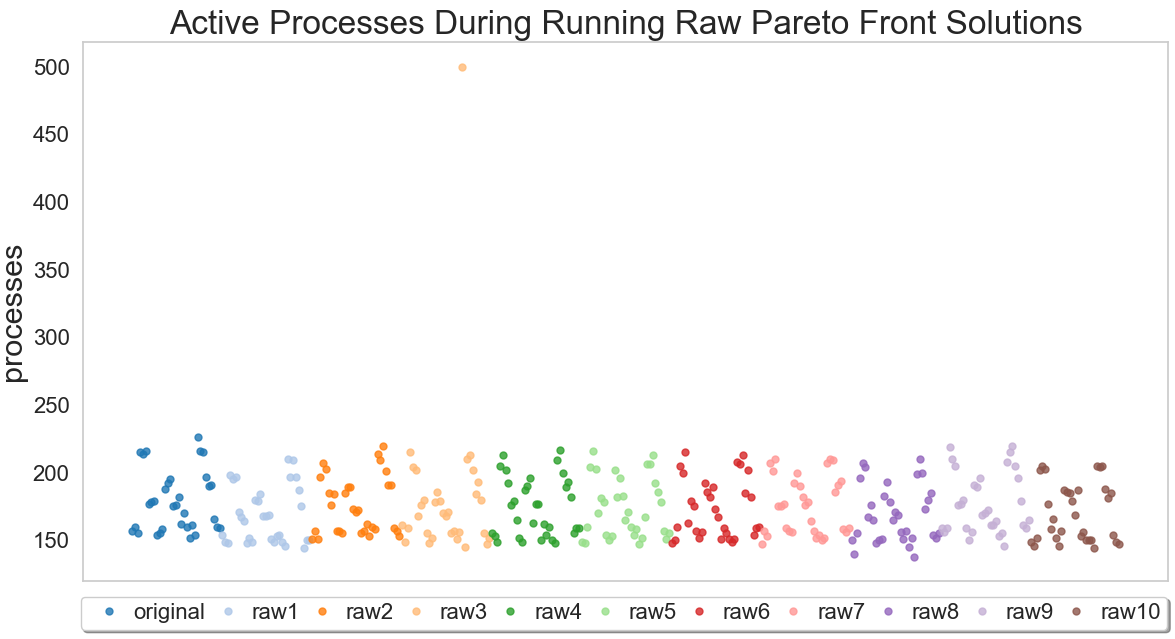}
  \caption{active processes}
  \label{activeProcessesRawExpByVariant-nexus9-2-M}
\end{subfigure}
\begin{subfigure}{.33\textwidth}
  \centering
  \includegraphics[width=.99\linewidth]{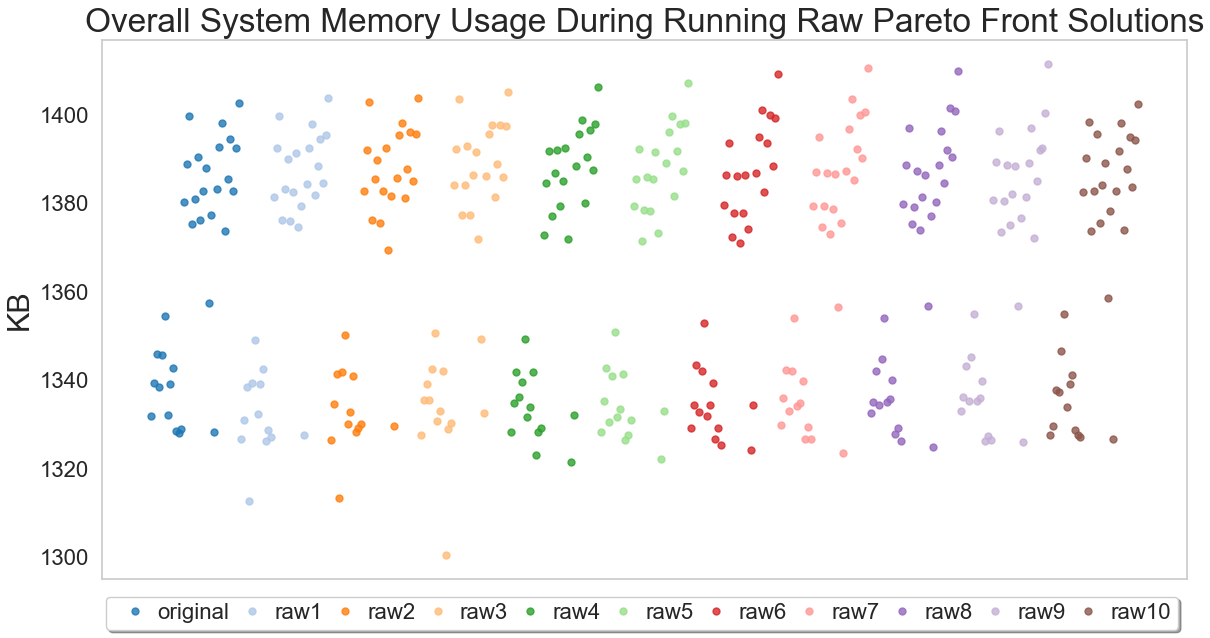}
  \caption{system memory usage}
  \label{overallSysMemoryUseRawExpByVariants-nexus9-2-M}
\end{subfigure}

\caption{Each solution in the Round-robin validation experiments after grouping them. Each solution's data is in different colour and the original configuration is in blue.}
\label{fig:R3results}
\end{figure*}

Now we show how \approachRRR helps mitigate the problem of changing system states by attempting to sample variants across similar sets of system states. 
\approachRRR works by sampling variants in a rotated round-robin style and then aggregating the measurements for each variant together in preparation for statistical comparisons. The beneficial effect of this on energy measurements can be seen in Figure~\ref{fig:violinRawRoundRobinN9-2-M} where distributions of energy measurements are compact and similar. Other system features show similar levels of consistency between variants. 

For example, Figure~\ref{scatterExecutionTimeReboundRawFronts-nexus9-2-M} shows the collected individual sample run-times for each variant

The dark blue samples represent the execution time of the original configuration. 
As can be seen, except for some outliers, there is remarkable 
consistency in the relative shape of the 
sequence of measurements of execution time with measurements for each variant generally trending up to a small peak and then dropping suddenly and then rising again. This consistency makes it relatively easy to distinguish between the underlying CPU utilisation of each variant.

This consistency of the set (and even sequence) of system states between variants under \approachRRR can be seen in the plots of active processes in  Figure~\ref{activeProcessesRawExpByVariant-nexus9-2-M}. With the exception of one outlier point, these sequences of active process readings have very similar shape and level.

Lastly, although the state of the overall system memory usage can not be fully controlled, our proposed approach ensured that each solution runs in similar overall main memory conditions. 
Figure~\ref{overallSysMemoryUseRawExpByVariants-nexus9-2-M} shows the overall system memory consumption during running each solution. Again, we can see that all configurations have been exposed fairly to the changing conditions. 

We have seen so far how much system state can change in hard-to-predict ways and how the \approachRRR scheduling approach can mitigate this problem to sample across similar system states. Next we compare \approachRRR in terms of precision and then in terms of sensitivity.


\section{Test Specificity}
\label{sec:specificity}
As mentioned previously \approachRRR is 
designed to maximise the similarity of the set of system states in which the energy used by program variants are sampled.
The intent of achieving this similarity is to produce datasets that improve the specificity and sensitivity of statistical tests run over that data. 
We demonstrate that \approachRRR scheduling produces energy readings that cater for more statistical specificity in our current experiments than the other approaches described above. We do this by showing that the other approaches are more prone to producing false positives. 

We start by sampling a corpus of repeated energy measurements from running the original Rebound variant (Rebound run 1000 times followed by a sample of energy use via the internal meter). These measurements span seven reboots on each platform\footnote{except Nexus 9A6 which exhibited caching problems in energy measurements.}. Because this corpus contains measurements from running the {\em{exact same variant}} a good validation schedule should sample these variants in an order such that statistical tests detect {\em{no difference}} between groups of samples. 

The seven reboots for each platforms supports seven samples each for $n=7$ simulated variants. Each scheduling approach will group these samples into different sets of seven samples for each of seven variants according to the schedule determined by each approach. We compare each of the variant groups, for each platform, for each approach using a two-tailed Wilcoxon rank-sum test with a threshold of $p\leq0.05$. 

\begin{figure*}[t]
\includegraphics[width=12cm]{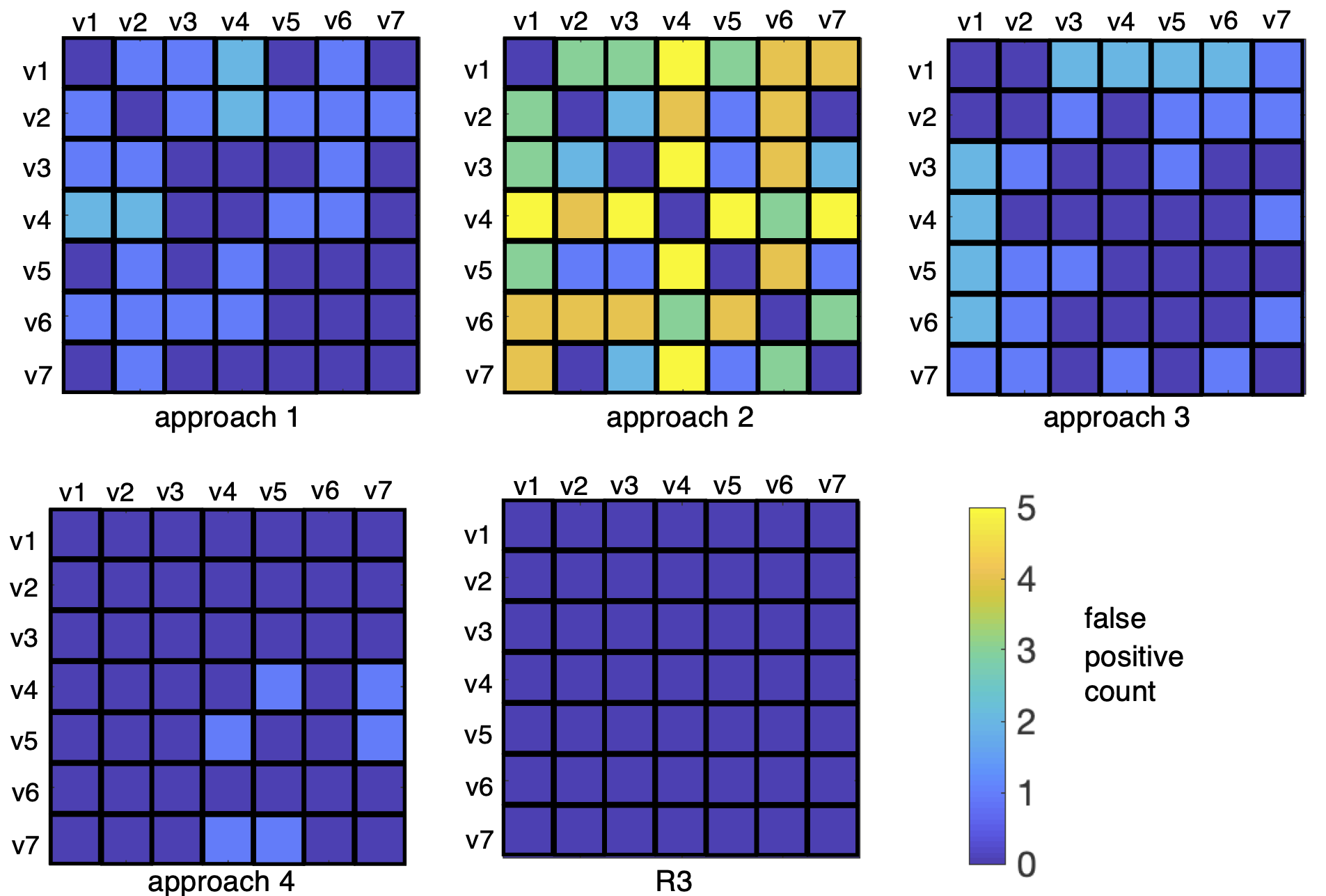}
\caption{False positive count for grouped variants across all platforms when data from the original Rebound variant is grouped into seven simulated program variants (v1,..,v7) by the five different sampling approaches. The only approach not to produce false positives is \approachRRR.}
\label{fig:specificity}
\end{figure*}

We aggregated the counts of false-positives ($p\leq0.05$) across all seven platforms and produced the matrices in Figure~\ref{fig:specificity}. 
It can be seen that the approach that produces the most false positives is \approachTwo this shows that rebooting has a significant impact on system state as reflected in energy measurements. \approachOne shows the effect of running variants far apart in time during a single charge cycle. The matrices for \approachThree and \approachFour indicate that interleaving has some success in capturing similar sets of system states. For \approachFour in particular, the only false positives are on the Sony Android 8 and 8.1 platforms. \approachRRR produces no false positives but produces a p-value close to the 0.05 threshold on the Sony Android 8.1 platform.

\section{Test Sensitivity of Validation Approaches}

\label{sec:comparison}

We now compare the test sensitivity of the validation approaches described in Section~\ref{sec:approaches} and our \approachRRR. 
As before we consider the Rebound variants, original, raw1,...,raw11 as the set of configurations to be validated.

To check how validation methods affect the statistical sensitivity of comparisons between each solution and the original configuration, we use an unpaired right-tailed Wilcoxon statistical rank-sum test. This non-parametric test does not make any assumptions about the distribution of the data-sets. Our null hypothesis is the original configuration's energy use samples are greater than the optimised solutions. The confidence level \textit{alpha} used is 0.05. 
To augment this comparison we also compute 
the Vargha and Delaney \^{A}\textsubscript{12} effect size to measure the approximate differences between the original configuration energy use and the optimised solutions. This measure is non-parametric and calculates the proportional difference between two data-sets~\cite{varghaDelanay}. It quantifies the difference in four ranges/thresholds for interpreting the effect size: 0.5 means no difference; up to 0.56 indicates a small difference; up to 0.64 indicates medium effect; over 0.71 is a large difference. This approach calculates the expected probability that solution 1 consumes less energy than solution 2. For example if \^{A}\textsubscript{12} = 0.8, then solution 1 is expected to consume less than solution 2 in 80\% of the time.

We use these statistical tests and measures for each approach on each 
platform in three ways. 
First, to report on the median effect size $median(es)$. 

Second, to measure how often we observe at least a ``medium'' effect in the direction of dominance, we count how often $es\geq0.64$. 
Third, we count the number of times the Wilcoxon rank-sum tests that indicate a significant difference at $p\leq0.05$. 

\begin{table*}[]
\caption{Performance comparison of different validation approaches on different platforms.
$es\geq0.64$ denotes the number of at least ``medium'' effects. 
$p\leq0.05$ denotes the number of Wilcoxon rank-sum tests that indicate a significant difference. The best values per device-OS configuration are highlighted in bold.}
\label{tab:rawPvalues}
\setlength{\tabcolsep}{2mm}
\renewcommand{\arraystretch}{1.0}
\begin{tabular}{c|ccc|ccc|ccc|ccc|ccc}
\toprule
 & \multicolumn{3}{c}{\approachOne} & \multicolumn{3}{c}{\approachTwo} & \multicolumn{3}{c}{\approachThree} & \multicolumn{3}{c}{\approachFour\ignore{approach 4 ABC*33}} & \multicolumn{3}{c}{\approachRRR} \\ 
 & 
\rotatebox[origin=c]{90}{median(es)} & 
\rotatebox[origin=c]{90}{$es\geq0.64$} & 
\rotatebox[origin=c]{90}{$p\leq0.05$} &
\rotatebox[origin=c]{90}{median(es)} & 
\rotatebox[origin=c]{90}{$es\geq0.64$} & 
\rotatebox[origin=c]{90}{$p\leq0.05$} &
\rotatebox[origin=c]{90}{median(es)} & 
\rotatebox[origin=c]{90}{$es\geq0.64$} & 
\rotatebox[origin=c]{90}{$p\leq0.05$} &
\rotatebox[origin=c]{90}{median(es)} & 
\rotatebox[origin=c]{90}{$es\geq0.64$} & 
\rotatebox[origin=c]{90}{$p\leq0.05$} &
\rotatebox[origin=c]{90}{median(es)} & 
\rotatebox[origin=c]{90}{$es\geq0.64$} & 
\rotatebox[origin=c]{90}{$p\leq0.05$} \\ \midrule
Nexus 9A6 & 0.57 & 5 & 5 & 0.38 & 1 & 2 & 0.73 & 7 & 7 & 0.65 & 6 & 7 & \textbf{0.82} & \textbf{10} & \textbf{10} \\
Nexus 9A7 & 0.42 & \textbf{3} & \textbf{3} & 0.27 & 1 & 1 & \textbf{0.50} & 1 & 1 &  &  &  &  &  &   \\
Nexus 6A6 & 0.50 & 5 & 5 & 0.17 & 1 & 1 & 0.83 & 9 & 9 & 0.75 & 8 & 9 & \textbf{0.87} & \textbf{10} & \textbf{10} \\
Nexus 6A7 & 0.12 & 0 & 0 & 0.57 & 4 & 3 & 0.60 & 5 & 5 & 0.50 & 0 & 0 & \textbf{0.82} & \textbf{10} & \textbf{10} \\
Moto GA6 & 0.38 & 1 & 2 & 0.72 & 7 & 8 & 0.72 & \textbf{9} & \textbf{9} & 0.70 & 8 & 8 & \textbf{0.9} & \textbf{9}  & \textbf{9} \\
Moto GA7 & 0.50 & 4 & 4 & 0.68 & 6 & 6 & 0.50 & 5 & 5 & 5 & 0 & 0 & \textbf{0.82} & \textbf{10} & \textbf{10} \\\bottomrule
\end{tabular}
\end{table*}

Table~\ref{tab:rawPvalues} reports the results of validating 11 software configurations using five validation approaches on six combinations of hardware and operating systems. {\color{black}Two blocks of data
are unavailable for \approachFour and \approachRRR for the Nexus 9 (Android 7) platform due to the device burning out under experimental load.}
It can be seen that, where data is available, \approachRRR is consistently better than other approaches on these measures. This indicates that, at least on these benchmarks and platforms \approachRRR exhibits higher sensitivity than other methods. We excluded the Sony XZ from our validation experiment since it was shown to have inaccurate energy readings in~\cite{bokhari2019mind}.

In addition, \approachOne and \approachTwo have the lowest performance among the approaches. The former does not take into consideration the variation in battery/voltage levels and their consequences on the system state. The latter, on the other hand, suffered from the random initial system states caused by rebooting.

Running solutions in round-robin fashion benefited \approachThree and \approachFour in most cases. In case of the former, reboots did not affect the collected samples, however, the fixed order did not expose all solutions to the same condition. For example, \emph{raw1} was always executed when the battery level was at 100\%. This issue has also impacted \approachFour. Both platforms however, performed well on Android 6 because solution run-times were twice as fast as on Android 7. This aligns with our findings presented in Section~\ref{singleExpOBD}.

\section{Conclusion}
\label{sec:conclusion}
Conventional means of validating software variants resulting from evolutionary processes are risky, because modern compute platforms exhibit a large number of system states beyond the control of a researcher. 
In this article, we have shown how a range of conditions -- such as system software states, memory consumption, and battery voltage -- can affect experimental results, which in turn may mislead the researchers. 
To mitigate the effects, we have proposed \approachRRR: it exercises the tests in a rotated-round-robin fashion, while accommodating the necessary regular restarting and recharging of the test platform. 
We have observed in a comparison with other validation approaches across multiple devices and operating systems that state changes have not misled our procedure, and that the measurable differences between configurations were the strongest for ours.


\bibliographystyle{acm}
\bibliography{library} 

\end{document}